\documentstyle[aps,epsfig]{revtex}

\def\deg{\ifmmode{^{\circ}}\else ${^{\circ}}$\fi}
\def\bi{\begin{itemize}}
\def\ei{\end{itemize}}
\def\ed{\end{document}}

\def\cf#1{\ifmmode{\cal #1}\else${\cal #1}$\fi}

\def\be{\begin{equation}}
\def\ee{\end{equation}}
\def\beas{\begin{eqnarray*}}
\def\eea{\end{eqnarray}}
\def\bea{\begin{eqnarray}}
\def\eeas{\end{eqnarray*}}

\def\gev{\ifmmode{\mbox{GeV}}\else GeV\fi}
\def\es{{\rm erg\ s}^{-1}}

\def\lf{L_{41}}

\def\emin{E_{\rm min}}
\def\emax{E_{\rm max}}
\def\deg{^{\circ}}
\begin{document}

\twocolumn
\renewcommand{\topfraction}{1.0}
\twocolumn[\hsize\textwidth\columnwidth\hsize\csname
@twocolumnfalse\endcsname

\title{A lower bound on the local extragalactic magnetic field}
\author{Luis A. Anchordoqui and Haim Goldberg}
\address{Department of
Physics, Northeastern University, Boston, MA 02115}

\maketitle

\begin{abstract}

Assuming that the hard $\gamma$-ray emission of Cen A is a result
of synchrotron radiation of ultra-relativistic electrons, we
derive a lower bound on the local extragalactic magnetic field,
$B> 10^{-8}$ G. This result is consistent with (and close to)
upper bounds on magnetic fields derived from consideration of
cosmic microwave background distortions and Faraday rotation
measurements.

\end{abstract}

\vskip2pc]

Increasing our  knowledge of the properties of extragalactic
magnetic fields (EGM) on scales between Mpc and the visible
horizon could significantly impact our understanding of phenomena
in cosmic ray physics (propagation, dissipation via synchrotron
radiation), cosmology (distortion of cosmic microwave background
(CMB)), and particle physics (early universe primordial
magnetogenesis during phase transitions). At present, surprisingly
little is actually known about EGM: there are some measurements
of diffuse radio emission from the bridge area between the Coma
and Abell superclusters \cite{kim,clarke}, which, under
assumptions of equipartition allows an estimate of ${\cal
O}(0.2-0.6)\,\mu$G for the magnetic field in this region. Such a
large field may possibly be understood if the bridge region lies
along a filament or sheet \cite{ryu}. Faraday rotation
measurements \cite{vallee,kronberg} have thus far served to set
upper bounds of ${\cal O}(10^{-9}-10^{-8})$ G on EGM on various
scales \cite{kronberg,olinto4}, as have the limits on distortion
of the CMB \cite{barrow,olinto3}. The Faraday rotation
measurements sample EGM of any origin out to quasar distances,
while the CMB analyses set limits on primordial magnetic fields
\cite{maartens}. We will discuss these bounds in some detail in
the concluding section, when comparing with our results.

In this paper, we will use the results of a set of cosmic ray
measurements taken over a ten-year period at the Sydney
University Giant Air-shower Recorder (SUGAR) \cite{winn1}, and
covering much of the southern sky, in order to set a {\em lower}
bound on the size of the $B$-field between Earth and the nearby
galaxy Centaurus A (Cen A), at a distance $d\approx 3.4$ Mpc. The
bound will be based on current models of dynamics in the ``hot spots'' of
radiogalaxies, and these assumptions will be  tested by future
observations of ultra-high energy cosmic ray (CR) and high energy
gamma ray fluxes from the direction of Cen A. We will initially
outline the dynamical picture concerning the generation of high
energy gamma rays. This picture leads to our Eq. (\ref{numrate}),
and then as a consequence to our lower bound on $B.$ We may
immediately note that we are aware of the crude angular and
energy resolutions of the SUGAR observations. Our analysis will
take into account these uncertainties.

Fanaroff-Riley II (FRII) galaxies \cite{fr} are the largest known
dissipative objects (non-thermal sources) in the universe.
Localized regions of intense synchrotron emission, so-called
``hot spots'', are observed within their lobes. These regions are
presumably produced when the bulk kinetic energy of the jets
ejected by a central active nucleus (supermassive black hole +
accretion disk) is reconverted into relativistic particles and
turbulent fields at a ``working surface'' in the head of the jets
\cite{blandford-rees}.
In what follows, we will adopt the first order Fermi
shock acceleration mechanism in hot spots of FRII glaxies, as discussed in Refs.
\cite{drury,rachen-b},
to account
for
particle acceleration to ultra high energies \cite{ostrowski}.
The applicability of this scenario to Cen A will be discussed in the
next paragraph.
The subtleties surrounding the conversion of particles' kinetic
energy into radiation provide ample material for discussion
\cite{bst,pic,felix}. The most popular mechanism to date relates
$\gamma$-ray emission to the development of electromagnetic
cascades triggered by secondary photomeson products that cool
instanstaneously via synchrotron radiation \cite{bst,pic}. The
characteristic single photon energy in synchrotron radiation
emitted by an electron is
\begin{equation}
E_\gamma = \left(\frac{3}{2}\right)^{1/2} \frac{h\,e\,E^2\,B}{2 \,\pi\, m_e^3 \,
c^5} \sim 5.4\, B_{\mu{\rm G}}\, E_{19}^{2} \, {\rm TeV}\ \ .\label{synch}
\end{equation}
For a proton this number  is $(m_p/m_e)^3 \sim 6 \times 10^9$
times smaller. Here, $B_{\mu{\rm G}}$ is the magnetic field in
units of $\mu$G and $E_{19} \equiv E/10^{19}$ eV. Thus, it is
evident that high energy gamma ray production through proton
synchrotron radiation requires very large (${\cal O}(100\ {\rm
G})$) magnetic fields.

We now discuss how the less luminous FRI galaxy Cen A fits
into this picture.
This radio-loud
source ($l \approx 310^\circ$, $b \approx 20^\circ$), identified
at optical frequencies with the galaxy NGC 5128, is the closest
example of the class of active galaxies \cite{israel}. Different
multi-wavelength studies have revealed a rather complex morphology:
it comprises a compact core, a jet also visible at $X$-ray
frequencies, a weak counterjet, two inner lobes, a kpc-scale
middle lobe, and two giant outer lobes. The jet would be
responsible for the formation of the northern inner and middle
lobes when interacting with the interstellar and intergalactic
medium, respectively. There appears to be a compact structure in
northern  lobe, at the extrapolated end of the jet. This
structure resembles the
hot spots like those existing at the extremities of FRII
galaxies. However, at Cen A it lies at the
side of the lobe rather than at the most distant northern edge,
and the brightness contrast (hot spot to lobe) is not as extreme
\cite{burns}.

In order to ascertain the capability of Cen A to accelerate
particles to ultra high energies, one first applies the Hillas
criterion \cite{hillas} for localizing the Fermi engine in space,
namely that the gyroradius $r_g= 110 E_{20}/B_{\mu{\rm G}}$ kpc
($E_{20}\equiv E/10^{20}$ eV) be less than the size of the
magnetic region.  Low resolution polarization measurements in the
region of the suspected hot spot give fields as high as $25\
\mu{\rm G}$ \cite{burns}. In certain of the regions where
measurements at both high and low resolution are available, the
$B$-field at high resolution can be twice that at low resolution.
The higher resolution can reveal  amplification in the post-shock
region \cite{ll}, yielding $B$-fields possibly as high as $50-60\
\mu{\rm G}$ \cite{burns,onecena}. The radio-visible size of the
hot spot can be directly measured from the large scale map of
Ref.\cite{junkes}, giving $R_{\rm HS}\simeq 2$ kpc. The actual
size can be larger by a factor $\sim 2$ because of uncertainties
in the angular projection of this region along the line of sight
\cite{counterjet}. If one assumes that the magnetic field of the
hot spot is confined to the visible region, then the limiting
energy is $\sim 2\times 10^{20}$ eV. However, it is plausible
that the shock structure in the hot spot extends beyond the
radio-visible region \cite{rachen-b,biermann2}.

In light of this, we apply to Cen A the
analyses of Refs. \cite{rachen-b,bst}, in which a limiting energy
is obtained by  balancing the characteristic
time scale for diffusive shock acceleration
\be
\tau_{\rm acc} \simeq \frac{40}{\pi}\, \frac{1}{c\,\beta_{\rm jet}^2}
\,\frac{1}{u}\,\left(\frac{E}{eB}\right)^{1/3}\, R^{-2/3}
\label{acc}
\ee
against the energy loss time scale
\be
\tau_{\rm loss} \simeq \frac{6\pi\ m_p^4\ c^3}{\sigma_T\ m_e^2\ B^2\ (1+Aa)}\ E^{-1}\ \ .
\ee
In the above, $\beta_{\rm jet}=$ jet velocity in units of $c,$ $u$ is the ratio of
turbulent to ambient magnetic energy density in the region of the shock
(of radius $R$),
$B$ is the total magnetic field strength, $a$ is the ratio of photon to magnetic
energy densities, $\sigma_T$ is the classical Thomson cross section, and $A$ is
a measure of
the relative strength of $\gamma p$ interactions against the synchrotron emission.
In Ref.\cite{rachen-b}, $A$ is estimated to be $\approx 200$ almost independent of source.
In Eq.(\ref{acc}) we assume that the turbulent component of the magnetic field
follows a Kolmogorov spectrum with spectral index $5/3.$
Equating these characteristic times yields a value for the maximum proton energy
\be
E_{20}=1.4\times 10^5\,\,B_{\mu{\rm G}}^{-5/4}\,\,\beta_{\rm jet}^{3/2}
\,\,u^{3/4}\,\,R_{\rm kpc}^{-1/2}\,\,(1+Aa)^{-3/4}\ \ ,
\label{ab}
\ee
where $R_{\rm kpc}\equiv R/1\ {\rm kpc}.$ One can estimate $u \sim 0.4$ from the
radio spectral index of synchrotron emission in the hot spot and the observed
degree of linear polarization in the same region \cite{jorgegus}.
The jet velocity is model dependent, and estimates  range from
$\sim 500$ km s$^{-1}$ to $0.99c$ \cite{burns}. For FRI galaxies, $a$ is expected to
be $\ll 1$ \cite{bst}, and  in our analysis we will sample a region of small $a.$
In Fig. 1 we plot the relation between $\beta_{\rm jet}$ and $a$ required to
attain various energies, for fiducial values $B=60\,\mu{\rm G},\ R=4$ kpc.
Since the range of values for $a$ and the jet velocity conform to
expected values, it is plausible that Cen A can accelerate particles to
energies $\agt 10^{20}$ eV.
\begin{figure}
\label{s0}
\begin{center}
\epsfig{file=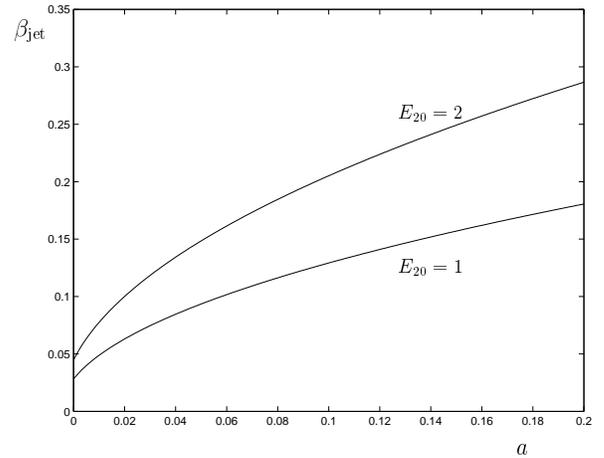,width=8.cm,clip=}
\hfill
\caption{Jet velocity as a function of parameter $a$ (defined in text),
for different proton energies.}
\end{center}
\end{figure}

Recent observations of the gamma ray flux for energies $>100$ MeV
(by the Energetic Gamma Ray Experiment Telescope (EGRET)
\cite{EGRET}) allow an estimate $L_{\gamma} \sim 10^{41}\ \es$
for the source \cite{aharonian}. This value of $L_{\gamma}$ is
consistent with an earlier observation of photons in the TeV-range
during a period of elevated activity \cite{g}, and is
considerably smaller than the estimated bolometric luminosity
$L_{\rm bol}\sim 10^{43}\es$\cite{israel}. Data across the entire
gamma ray bandwidth of Cen A is given in Ref.\cite{steinle},
reaching energies as high as 150 TeV \cite{150}. Data at this energy await
confirmation. For  values of $B$ in the $\mu$G range,
substantial proton synchrotron
cooling is suppressed, allowing production of high energy
electrons through photomeson processes. The average energy of
synchrotron photons scales as $\overline{E}_\gamma \simeq 0.29
E_\gamma$ \cite{gs}. With this in mind, it is straightforward to
see that to account for TeV photons Cen A should harbor a
population of ultra-relativistic electrons with $E \sim 6 \times
10^{18}$ eV. We further note that this would require the presence
of protons with energies between one and two orders of magnitude
larger, since the electrons are produced as secondaries \cite{gopal}.

There are plausible physical arguments \cite{pic,waxman} as well
as some observational reasons \cite{steve-richard} to believe
that when proton acceleration is being limited by energy losses,
the CR luminosity $L_{\rm CR}\approx L_{\gamma}$. In the spirit
of \cite{agw}, we introduce $\epsilon$, the efficiency of ultra
high energy CR production compared to high energy $\gamma$
production---from the above, we expect $\epsilon\simeq 1.$ Using
equal power per decade over the interval $(\emin,\ \emax)$, we
estimate a source luminosity
\begin{equation}
\frac{E^2 \,dN^{p+n}_0}{dE\,dt} \, \approx \frac{6.3\,\epsilon\lf
\,10^{52} {\rm eV/s}}{\ln(\emax/\emin)} \label{2}
\end{equation}
where $\lf
\equiv$ luminosity of Cen A$/10^{41}\es$ and the subscript ``0'' refers
to quantities at the source.

For a relatively close source like Cen A, one can neglect
interactions of cosmic ray protons with the universal background
radiations. Consequently, the shape of the spectrum would be
unmodified \cite{ac}. In order to discuss directionality, we first consider
the case where there is {\em no} intergalactic magnetic field. Then {\it en route}
to us the protons suffer no deflections and no spectral distortions.
With the source luminosity (\ref{2}), the counting rate at Earth in an energy bin
$(E_1, E_2)$ for a detector of
area $S$ due to Cen A would be
\begin{eqnarray}
\left.\frac{dN}{dt}\right|_{B=0} & = &  \frac{S }{4 \pi d^2}
\,\,\int_{E_1}^{E_2}
    \frac{E^2\,dN_0}{dE\,dt} \, \frac{dE}{E^2} \nonumber \\
& = & 14.3\,\,\frac{(S/100\ {\rm km}^2)\,\, \epsilon
\lf}{\ln(\emax/
\emin)}\nonumber\\
&&\,\, \times\,\left(\frac{1}{E_{1,20}}-\frac{1}{E_{2,20}}\right)
\,\,\,\, {\rm events}/{\rm yr} \label{rate}.
\end{eqnarray}
(Here $E_{1,20}=E_1/10^{20}\,{\rm eV}$, etc.)
This would all be concentrated in a cone of half-angle $\theta_{\rm res},$
the observational resolution, about the direction of Cen A.

As mentioned previously, the only existing measurements of
ultrahigh energy cosmic rays arriving from  directions which
include Cen A were carried out at SUGAR during a ten year period
1968-1979. The array was located at $30\deg 32'$ S, $149\deg 46'$
E, and presented a total area of 100 km$^2$. Shown as asterisks
in Fig. 2, as given in Ref. \cite{winn1}, are the arrival
directions of 80 events with energies above $4\times 10^{19}$ eV.
The direction of Cen A is indicated with a five-pointed star, and
the experimental 1$\sigma$ uncertainties in arrival directions of
events near the Cen A direction are indicated by dashed lines
\cite{res}. The solid line ovals surrounding Cen A indicate
regions within $10\deg$ and $25\deg,$ respectively, of Cen A, and
their significance will be explained shortly.

We now obtain the $B=0$ event rate expected at SUGAR in the
direction of Cen A, as predicted by  Eq.(\ref{rate}). The energy
bin is appropriate to the data of \cite{winn1}, $E_{1,20}=0.4,\
E_{2,20}=2.$ A conservative lower bound on the energy of protons
which are progenitors of high energy gamma radiation is $E_{\rm
min}=1\times 10^{19}$ eV. It also reasonable to take for the
upper bound $E_{\rm max}=4\times 10^{20}$ eV \cite{emax}. With this input, we
obtain from Eq.(\ref{rate})
\begin{equation}
\left.\frac{dN}{dt}\right|_{B=0}  \agt  8\ \epsilon\ \lf\,\,\,
{\rm events/year}\ \ . \label{numrate}
\end{equation}
In the preceding discussion we have indicated that under
plausible conditions may expect $\epsilon\simeq 1.$ However, even
if $\epsilon\lf$ is as small as 0.1, we may, in the absence of a
magnetic field, expect 8 events in 10 years from the direction of
Cen A.

The cosmic ray orbits undergo bending in both the Galactic and
extragalactic  magnetic fields. The magnetic deflection of
protons in the Galactic disk has been studied in detail in
\cite{todor}. This analysis includes two extreme options for the
behavior of the field, reflecting the different symmetries with
respect to field reversals in the $r$- and $z$-directions. The
$B$-field has a $1/r$ behavior, with deviations calculated out to
20 kpc from the Galactic center. The r.m.s. deviation (averaged
over arrival direction), for an energy of $4\times 10^{19}$ eV,
varies from $7.9\deg$ to $10.5\deg$ in going between the two
models. This deviation shows an approximately linear decrease
with increasing energy. All the events which we will consider
have energies $>4\times 10^{19}$ eV, so that conservatively, we
allow an uncertainty of $10\deg$ in arrival direction due to
deviation in the Galactic magnetic field. Events within the inner
oval of Fig. 2 represent events which could have originated from
Cen A, and have suffered deflection only in the Galactic field.
It is clear that at most 1 event can be categorized in this
manner, so that in order to have at least 8 events (see previous
paragraph), some amount of extragalactic field is necessary.

We treat the extragalactic deviation in a standard manner
\cite{we}. With $B_{{\rm
nG}}=B/10^{-9} {\rm G},$ the Larmor radius of a particle in this
field is $r_{\rm L} \simeq 10^2$ Mpc $E_{20}/B_{\rm nG}.$ If this
is sizeably larger than the coherence length of the magnetic field
$\ell_{\rm coh}$, the typical deflection angle from the direction
of the source, located at a distance $d$, can be estimated
assuming that the particle makes a random walk in the magnetic
field \cite{we}
\begin{equation}
\theta(E) \simeq 0.54^\circ\, \left(\frac{d}{1\ {\rm Mpc}}\right)^{1/2}\,
\left(\frac{\ell_{\rm coh}}{1 \,\, {\rm Mpc}} \right)^{1/2} \,
\frac{B_{{\rm nG}}}{E_{20}}\ \ .\label{random}
\end{equation}
Thus, in the case that $\theta(E)\ll 1,$ we expect that  the
counting rate (\ref{rate}) would be spread out over a cone of
half-angle
\begin{equation}
\overline{\theta}(E)\simeq \left(\theta(E)^2+\theta_{\rm res}^2\right)^{1/2}\ \ .
\label{thetavg}
\end{equation}

We now draw a second oval in Fig. 2 to include the least number of
events compatible with the expectation from Cen A. This second
oval contains (or partially contains) 7 events within $25\deg$ of
Cen A. From our previous discussion of the Galactic effects, this
leads us to conclude that {\em cosmic rays with $E>4\times
10^{19}$ eV experience a deviation of at least $15\deg$ in
extragalactic magnetic fields during their transit from {\rm Cen
A} to {\rm Earth}}.

In conjunction with Eqs.(\ref{random}) and (\ref{thetavg}) we
utilize this conclusion  to obtain a lower bound on the $B$-field
between Earth and Cen A. The average energy $\overline E$ of the
cosmic ray events above $4\times 10^{19}$ eV, as reported in
\cite{winn1}, is $6.3\times 10^{19}$ eV. Requiring
$\overline\theta(\overline E) > 15\deg,$ with $\theta_{\rm res} =
3\deg$ and $d=3.4$ Mpc, we obtain from (\ref{random}) and
(\ref{thetavg})
\be B>9.5\times 10^{-9}\,\left(\ell_{\rm coh}/1\,
{\rm Mpc}\right)^{-1/2}\ {\rm G}\ .
\label{bounda}
\ee Since the
coherent length is most likely $< 1$ Mpc, we obtain the bound
stated in the abstract,
\be B > 1.0\times 10^{-8}\ {\rm G}\ \ .
\label{boundb}
\ee

\begin{figure}
\label{s1}
\begin{center}
\epsfig{file=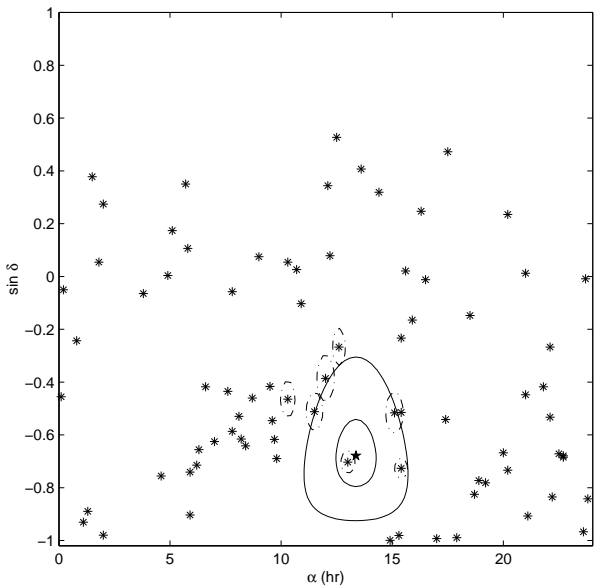,width=8.cm,clip=}
\hfill
\caption{The
nominal arrival directions ($\alpha =$ right ascension, $\delta =$
declination) of SUGAR events with energies above $4 \times
10^{19}$ eV. Also shown in solid lines are contour maps
indicating the circular areas of the celestial sphere centered at
Cen A (indicated by $\star$) with $10^\circ$ and $25^\circ$
radii. The dashed lines surrounding several of the events
indicate the angular resolution of the experiment.}
\end{center}
\end{figure}

Our analysis has utilized the SUGAR data with energies assigned
according to the Hillas model \cite{hillas}. With this
prescription, the energy spectrum agrees with that of other
experiments \cite{winn1}. An alternate method \cite{sydney} which
underestimates the energy by a factor of more than 2 leads to a
spectrum which does not show this agreement \cite{winn1}. The
lower energy would have two effects in our analysis: $(i)$ the
expected event rate (\ref{numrate}) would double because of the
smaller lower limit on the energy bin $(ii)$ the deviation in the
Galactic magnetic field would roughly double \cite{todor},
leading to an increase in the size of the inner oval in Fig. 2.
Following the same procedure as previously, and still utilizing
the conservative bound $\epsilon\lf\ge 0.1,$ we would find that
$B>5$ nG. However, because there is  general consensus
\cite{kewley} that the spectral evidence strongly favors the
Hillas energy normalization, we maintain our bound as $B>$ 10 nG
\cite{lower}.

The limit (\ref{boundb}) has an immediate implication with respect
to lower energy cosmic rays:  from Eq.(\ref{random}), one can
conclude that if $B>$ 10 nG all directionality is lost for protons
with energies below $\sim 2\times 10^{19}$ eV. This is consistent
with the absence of any anisotropy in the events observed at
SUGAR at these lower energies \cite{winn2}.

We turn to discuss our result in light of  considerations which
impose {\em upper} bounds on magnetic fields. (1) An analysis of
He$^4$ production provides a weak bound $B_0<10^{-6}$ G on a
homogeneous primordial field \cite{olinto1}; however, such fields
are largely dissipated prior to nucleosynthesis \cite{olinto2}.
(2) From limits on distortions of the cosmic microwave background
(CMB), there is an upper bound $B_0< 3.4\times 10^{-9}$ G for
fields homogeneous on present horizon scales \cite{barrow}. Our
lower bound (\ref{boundb}) exceeds this, so that we may conclude
that the magnetic field from here to Cen A contains components in
excess of a homogeneous relic background field. (3) The analysis
of CMB distortions extended to inhomogeneous fields provides a
larger bound $B_0< 3\times 10^{-8}$ G on scales between 400 pc and
0.6 Mpc \cite{olinto3}. This lies above our  limit, so that the
local EGM is consistent with being of primordial origin. (4) The
absence of a positive signal in Faraday rotation measurements on
QSO's \cite{kronberg} provides upper limits on magnetic fields
(of any origin) as a function of reversal scale. These bounds
depend significantly on assumptions about the electron density
profile as a function of red shift $z.$ When electron densities
follow that of the Lyman-$\alpha$ forest, the average magnitude
of the magnetic field receives an upper limit of $B<10^{-9}$ G for
reversals on the scale of the horizon, and $B<10^{-8}$ G for
reversal scales on the order of 1 Mpc \cite{olinto4}.  The latter
{\em upper} bound is roughly coincident with our {\em lower} limit
in our galactic neighborhood. Local fluctuations in electron
densities on scales of 1-30 Mpc can lead to very large
concommitant fluctuations in the corresponding magnetic field
\cite{ryu}, so that the bound in \cite{olinto4} should be read as
averaged over many reversal cells between Earth and the light
sources at distances out to $z=2.5.$ If it should happen that the
fluctuations in the magnitude of $B$ in the Mpc cells out to the
horizon are of the same order as $B_{\rm avg}$ itself, then our
result can  imply that the {\em average value of the
intergalactic magnetic field is of the order of $10^{-8}$ G.}
Because of the $B^2$ dependence of synchrotron radiation loss by
secondary electrons, a field of this strength can have important
implication on the development of electromagnetic cascades. Such
cascades  are characteristically associated with the decay of
supermassive relic particles or topological defects. An average
magnetic field of $10^{-8}$ G will impose strong constraints on
``top-down'' models \cite{protheroe}.

The SUGAR observations used in this analysis were recorded more
than twenty years ago. Both resolution and statistics will be
vastly improved with data to be available from facilities which
will observe the southern sky. These are now coming on line (Auger
\cite{auger}), or are expected to (EUSO/OWL \cite{eusowl}) in the
not-too-distant future. These data will significantly enhance our
knowledge of magnetic field strengths in the extragalactic
neighborhood of the Milky Way.

 \hfill

We would like to thank Peter Biermann for sharing with us his
expertise  about magnetic fields in shock environment.  This work
was partially supported by CONICET (LAA), and the National Science
Foundation (HG).


\begin{thebibliography}{99}
\bibitem{counterjet} For example, an explanation of  the apparent
absence of a counterjet in Cen A via relativistic beaming
suggests that the angle of the visible jet axis with respect to
the line of sight is at most 36$^{\circ} \cite{burns},$ which
could lead to a doubling of the hot spot radius. It should be
remarked that for a distance of 3.4 Mpc, the extent of the entire
source has a reasonable size even with this small angle.
\bibitem{gopal} Consecutive factor of 2 energy losses occur in the
processes $p\gamma\rightarrow N\pi^0,\ \pi^0\rightarrow
\gamma\gamma,\ \gamma\rightarrow e^+e^-.$ Eq.(\ref{synch}) then
implies proton energies of $\sim 10^{20}$ eV for 100 TeV photons.
\bibitem{res}This is given in Ref. \cite{winn1} as $\theta_{\rm
res}=3^{\circ}\ \sec \theta_{\rm zenith}.$ The several events at
declinations $\delta$ near $-90\deg$ have small errors in the
$\delta$ direction, and do not intersect the oval regions.
\bibitem{emax} Our results are insensitive to lowering $E_{\rm max}$ to
$2\times 10^{20}$ eV, or lower.
\bibitem{lower} A lower cutoff on the maximum attainable energy
would increase the expected event rate, and would decrease the
possible observed events, leaving the bound unchanged.
\end{thebibliography}
\end{document}